\documentclass[twoside]{ilcws07}
\usepackage[latin1]{inputenc}
\usepackage[dvips]{graphicx,epsfig,color}
\usepackage{wrapfig,rotating}
\usepackage{amssymb,amsmath,array}

\begin{document}
\title{
%%%%   Paper title goes here  %%%%%%%%%%%%%%
Simulation of LiCAS Error Propagation} %% 
%***********************************************************************
% AUTHORS INFORMATION AREA
%***********************************************************************
\author{G.\ Grzelak$^1$, 
        A.\ Reichold$^2$,
        J.\ Dale$^2$,
        M.\ Dawson$^2$,
        J.\ Green$^2$, \\
        Y.\ Han$^2$,
        M.\ Jones$^2$,
        G.\ Moss$^2$,
        B.\ Ottewell$^2$,
        R.\ Wastie$^2$, \\
        D.\ K\"amptner$^3$,
        J.\ Prenting$^3$,
        M.\ Schl\"osser$^3$
% Optional short acknowledgement: remove next line if non-needed
%% \thanks{This is an optional funding source acknowledgement.}
% DO NOT MODIFY THE FOLLOWING '\vspace' ARGUMENT
\vspace{.3cm}\\
% Addresses and institutions (remove "1- " in case of a single institution)
1- University of Warsaw - Institute of Experimental Physics, \\
ul.\ Ho\.za 69, 00-681 Warsaw - Poland
\vspace{.1cm}\\
2- University of Oxford - Department of Physics \\
Denys Wilkinson Building, Keble Road, Oxford OX1 3RH - United Kingdom
\vspace{.1cm}\\
3- Deutsches Elektronen Synchrotron, DESY, \\
   Notkestrasse 85, 22607 Hamburg - Germany
}
%%***********************************************************************
% END OF AUTHORS INFORMATION AREA
%***********************************************************************

\maketitle

\begin{abstract}
  Linear Collider Alignment and Survey (LiCAS) R\&D group is proposing
a novel automated metrology instrument dedicated to align
and monitor the mechanical stability of a future linear high energy $e^{+}e^{-}$ collider.
LiCAS uses Laser Straightness Monitors (LSM) and Frequency Scanning 
Interferometry (FSI) \cite{Coe,Fox-Murphy}
for straightness and absolute distance measurements, respectively.
  This paper presents detailed simulations of a LiCAS system operating 
inside a Rapid Tunnel Reference Surveyor (RTRS train).
  With the proposed design it is feasible to achieve the required vertical 
accuracy of the order of $\mathcal{O}(200)\,\mu m$ over $600\,m$ tunnel
sections meeting the specification for the TESLA collider~\cite{TESLA}.
\end{abstract}

\section{Principle of the LICAS-RTRS train operation}

\begin{figure}
\begin{center}
\begin{tabular}{c}
\includegraphics[width=12cm]{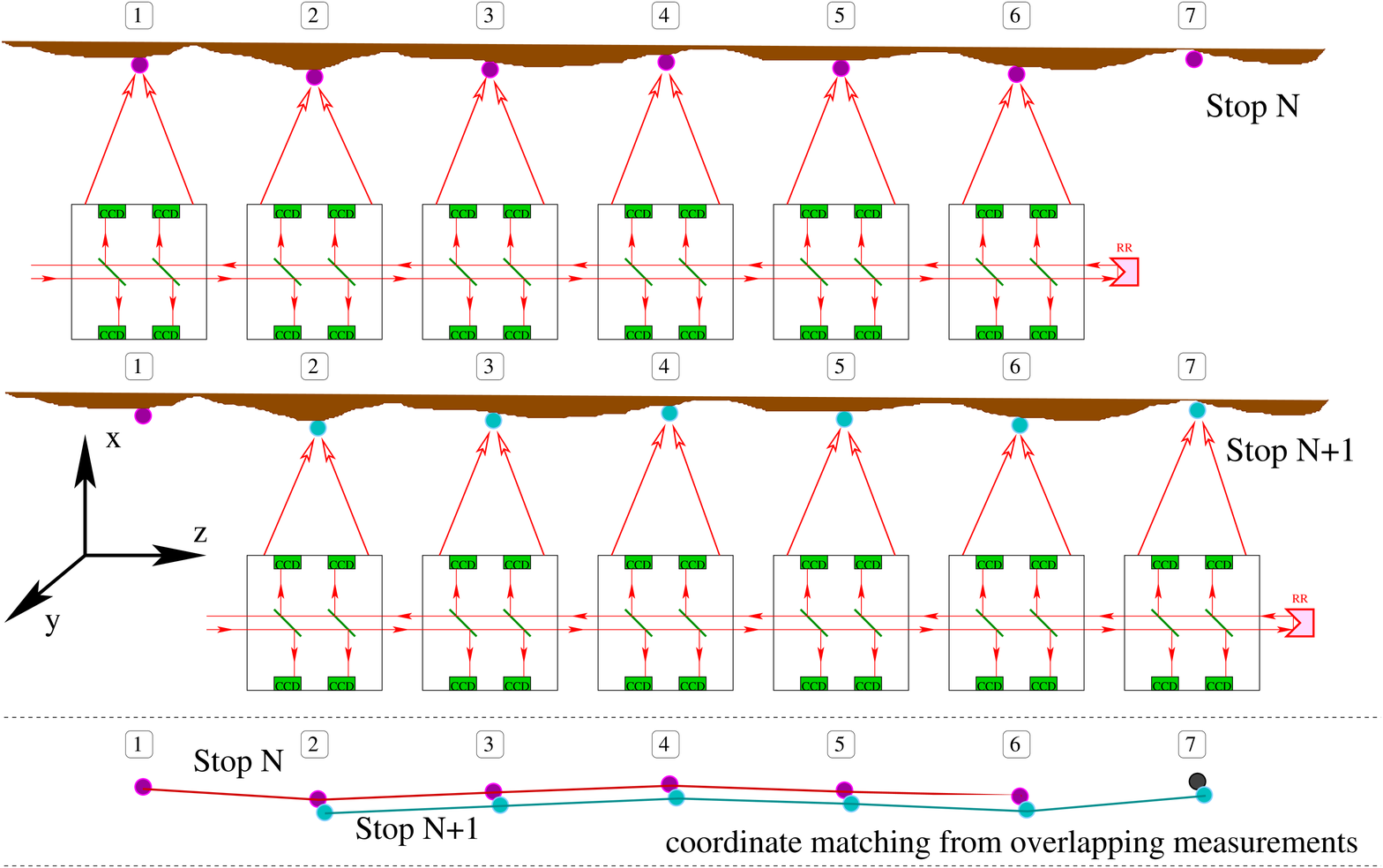}
\end{tabular}
\end{center}
\caption[Principle of the LiCAS train operation.]
{Principle of the LiCAS train operation.
Top view of the two train stops along the accelerator tunnel are presented.
For detailed description see text.}
\label{TRAIN-PRINC}
\end{figure}

In figure \ref{TRAIN-PRINC} the schematic view of the LiCAS train
operating in the accelerator tunnel is presented. The train is composed
of 6 cars, the distance between the centres of neighbouring cars is $\sim4.5\,m$.
Each car is equipped with 4 CCD cameras and two beam splitters (BS) constituting
the straightness monitor. 
The straightness monitor measures the transverse translation
($T_{x},T_{y}$) and transverse rotation ($R_{x},R_{y}$) with respect to a
$z$ axis defined by the laser beam passing through all cars in a vacuum pipe.
The laser beam is reflected back using the retro-reflector (RR) located in the last car,
illuminating the upper CCD cameras of the straightness monitors.
6 FSI lines placed in the same vacuum pipe between each pair of cars are responsible for 
the distance 
measurement along the $z$ axis ($T_{z}$). In addition a clinometer located on each car
provides a measurement of rotation around the $z$ axis ($R_{z}$).
%------------------------------------------------------------------------------
%------------------------------------------------------------------------------
When the train stops in front of the wall markers it firstly measures the
relative position and rotation of all cars with respect to the first car.
This defines the local reference frame of the train in which the location
of the wall mounted reference markers are measured next. This procedure is
repeated for each train stop. Each marker is measured up to 6 times.
Finally the coordinates of each marker, expressed in the local train frames
are transformed to the frame of the first train (the global frame) 
by fitting them to each other under the constraint that wall markers have
not moved during the entire measurement.

\section{Opto-geometrical model of the LICAS-RTRS train}

In order to study the expected precision on the position reconstruction
of the tunnel reference markers a simulation of the
LiCAS survey train was performed. To describe the sensing parts of the
train the {\sc Simulgeo} \cite{Brunel} package was used which allows for 
modelling of the opto-geometrical systems.
This software is also capable of performing the full error propagation including 
correlations between various sub-components linked via common mechanical supports.

\subsection{Results of the train simulations}

Presented results were obtained assuming the intrinsic resolution of the
CCD cameras and FSI lines equal to $\sigma_{CCD}=\sigma_{FSI}=1\,\mu m$.
The assumed precision of the clinometer was $\sigma_{tilt}=1\,\mu rad$.
The simulation was performed under the assumption that all
calibration constants (positions and rotations of CCD cameras, beam
splitters, FSI light sources and retro-reflectors) are known to
the accuracy of $\sigma_{pos}=1\,\mu m$ for positions and 
$\sigma_{ang}=1\,\mu rad$ for angles. 

% $$$ Nevertheless the ``$1\,\mu m$''/``$1\,\mu rad$''
% simulations can be regarded as the reference to be compared to results
% obtained with more realistic values of calibration parameter errors 
% in the future.

The long-distance operation of the train inside the
accelerator tunnel was simulated by a set of many identical
trains displaced by $4.5\,m$ (distance between stops), each pair of them coupled 
via 5 overlapping wall markers.
{\sc Simulgeo} calculations provide very precise results 
(taking into account correlations between subcomponents of the system) 
based on the exact opto-geometrical model of the survey procedure.
However, from the numerical point of view, such an approach,
manipulating large matrices, is very time and memory consuming. 
The 20 train stop results ($90\,m$ tunnel section) were obtained 
after 34 hours of CPU time using
$1\,GB$ RAM memory on a $2\,GHz$ machine (the rank of the used matrix
was of the order of $10\,000$). The numerical complexity
of these calculations scale like $N^2$, where $N$ is the number
of involved coordinates. The simulation of the full $600\,m$ tunnel
section would require more then 7 weeks of CPU time.

\subsection{Random walk model}

To overcome the above mentioned limitations a simplified analytical formula 
inspired by a random walk model was derived to extrapolate 
the {\sc Simulgeo} predictions over long tunnel sections:
\begin{equation}
\sigma_{xy,n} = \sqrt{l^2\sigma^2_\alpha \frac{n(n+1)(2n+1)}{6}
              + \sigma^2_{xy} \frac{n(n+1)}{2}}, \,\,\,\,\,
\sigma_{z,n}  = \sqrt{\sigma^2_z \frac{n(n+1)}{2}}
\label{RANDOM-WALK-FORMULA}
\end{equation}
where $n$ is the wall marker number, $l$ is the effective 
length of the ruler (here: distance between cars),
and the corresponding errors are the parameters of the
random walk: $\sigma_\alpha$ is the angular error,
$\sigma_{xy}$ are the transverse errors
and $\sigma_{z}$ is the longitudinal error.
In this approach the procedure of accelerator alignment
resembles the construction of a long straight line using short ruler. 
The overall error is
a convolution of the precision of the ruler and the precision of
the placement of the ruler with respect to the previous measurement.
The asymptotic behaviour of the formulae from equation 
no.\ \ref{RANDOM-WALK-FORMULA} is:
$\sigma_{xy,n} \sim n^{\frac{3}{2}}$, and $\sigma_{z,n}  \sim n$.
This fast growth of errors (especially for transverse directions) 
is a consequence of the fact that the errors are highly correlated
and the precision of the $n^{th}$ element depends on the precision
of all previous points.
%------------------------------------------------------------------------------
\begin{figure}
\begin{center}
\begin{tabular}{c}
\includegraphics[width=4.6cm]{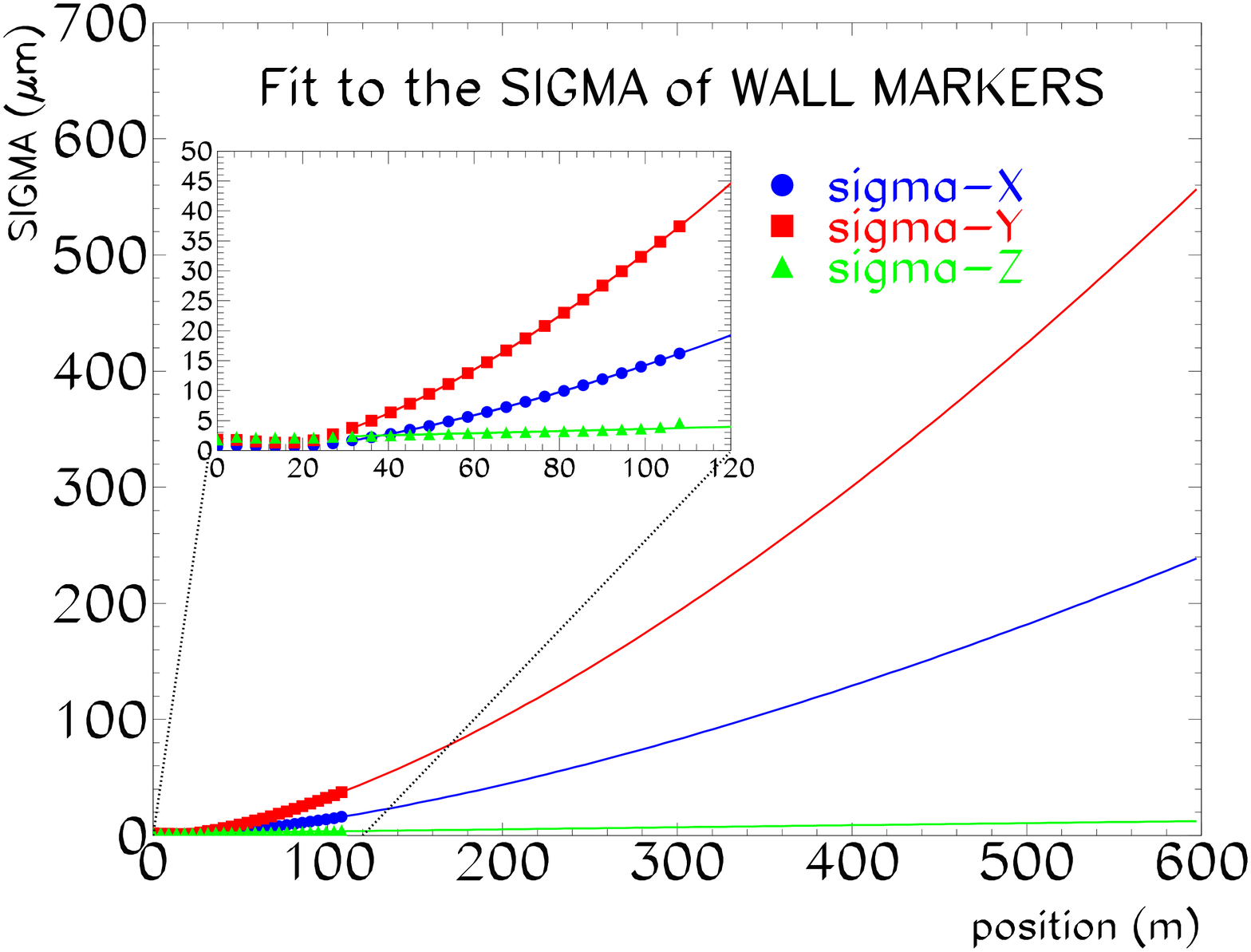}
\includegraphics[width=4.6cm]{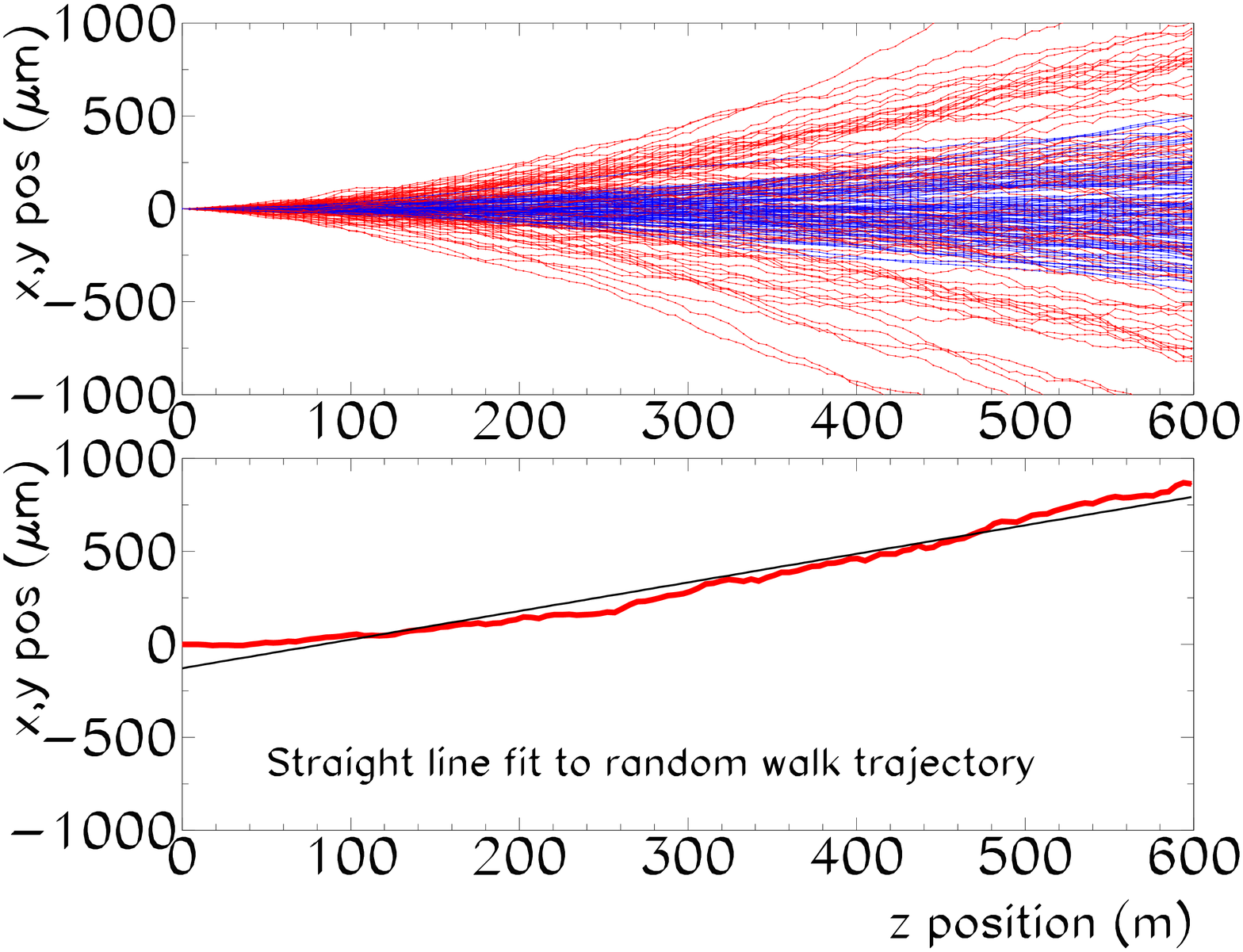}
\includegraphics[width=4.6cm]{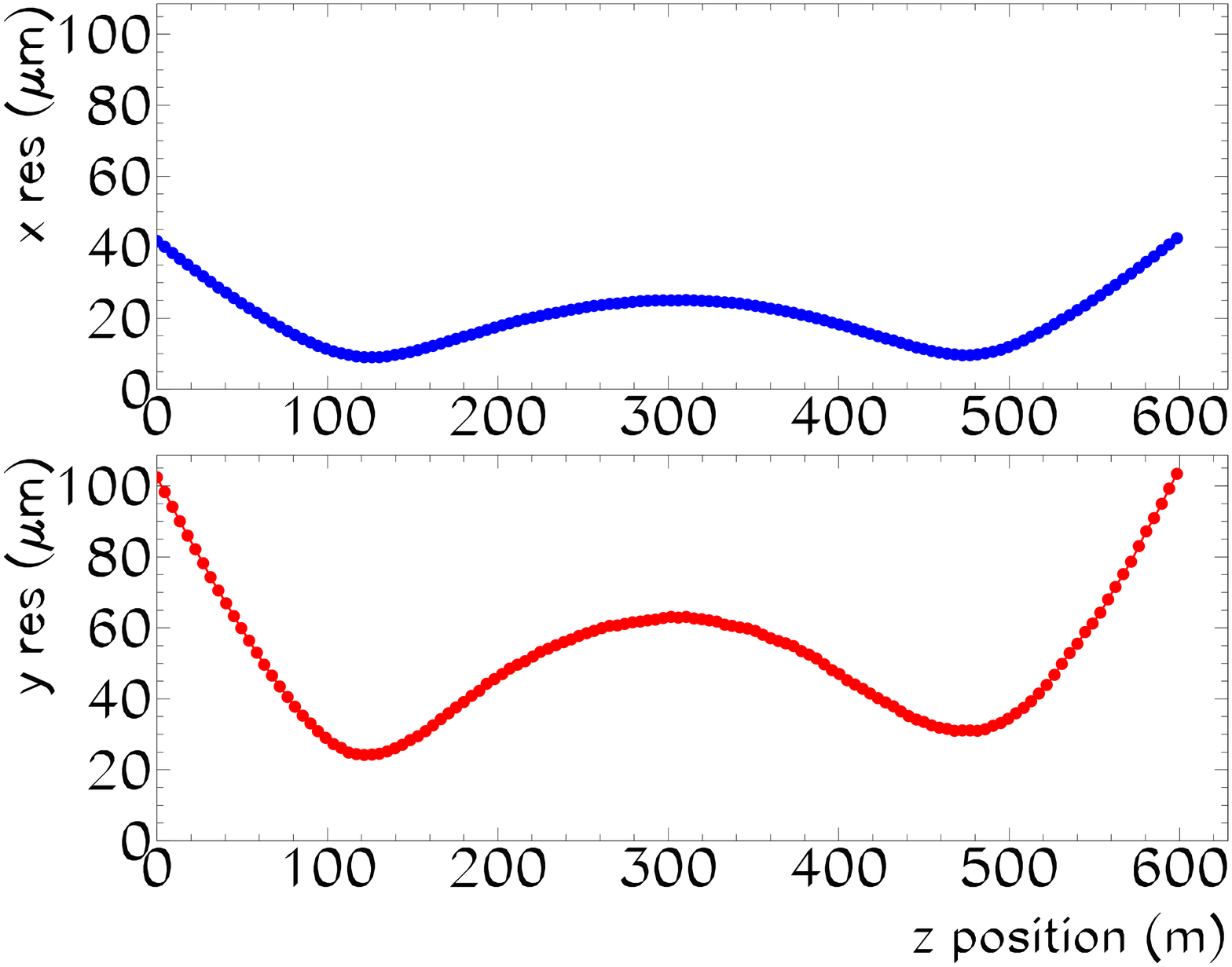}
\end{tabular}
\end{center}
\caption[Results of the LiCAS train simulation]
{LiCAS train simulation (left plot): insertion contains the 
results of the exact simulation obtained with the {\sc Simulgeo},
extrapolated on the main plot to the $600\,m$ tunnel section using
the formula from the random walk model.
Examples of the random walk trajectories (upper middle plot)
and straight line fit to selected trajectory (lower middle plot).
RMS of the residua distribution from the straight line fit
to the random walk trajectories: horizontal (upper right plot) and vertical
direction (lower right plot).}
\label{SIMULGEO-AND-RANDOM-WALK}
\end{figure} 
%------------------------------------------------------------------------------
Formulae \ref{RANDOM-WALK-FORMULA} were fitted to the {\sc Simulgeo}
points determining $\sigma_\alpha$, $\sigma_{xy}$, $\sigma_{z}$
and then extrapolated over a $600\,m$ tunnel section 
(fig.\ \ref{SIMULGEO-AND-RANDOM-WALK} left plot).
The obtained predictions refer to the precision of the placement of the $n^{th}$
accelerator component with respect to the first one. However this is not the
ultimate measure of the quality of the accelerator alignment. The relevant
parameter is the mean deviation of each component from the ideal straight line 
which can be expected from the above procedure. 
To obtain the final prediction on the deviation of the alignment from 
the straight line a series of random walk trajectories was generated using the
parameters fitted to the {\sc Simulgeo} points 
(fig.\ \ref{SIMULGEO-AND-RANDOM-WALK} middle column). 
A straight line was fitted to each trajectory and the corresponding
residua were calculated.
The extracted RMS values of the residua distributions for each marker 
along $600\,m$ provide the measure of the accuracy of the whole procedure. 
Because of high correlation between errors for $n^{th}$ and $(n+1)^{th}$ marker
the generated trajectories exhibit much smaller oscillations that
would be expected from completely random process.
Figure \ref{SIMULGEO-AND-RANDOM-WALK} (right column)
summarises the results obtained in this 
analysis demonstrating that the vertical precision of the order 
of $\mathcal{O}(100\,\mu m)$ over $600\,m$ is feasible.

\subsection{Spectral analysis of the alignment trajectories}

\begin{figure}
\begin{center}
\begin{tabular}{c}
\includegraphics[width=8cm]{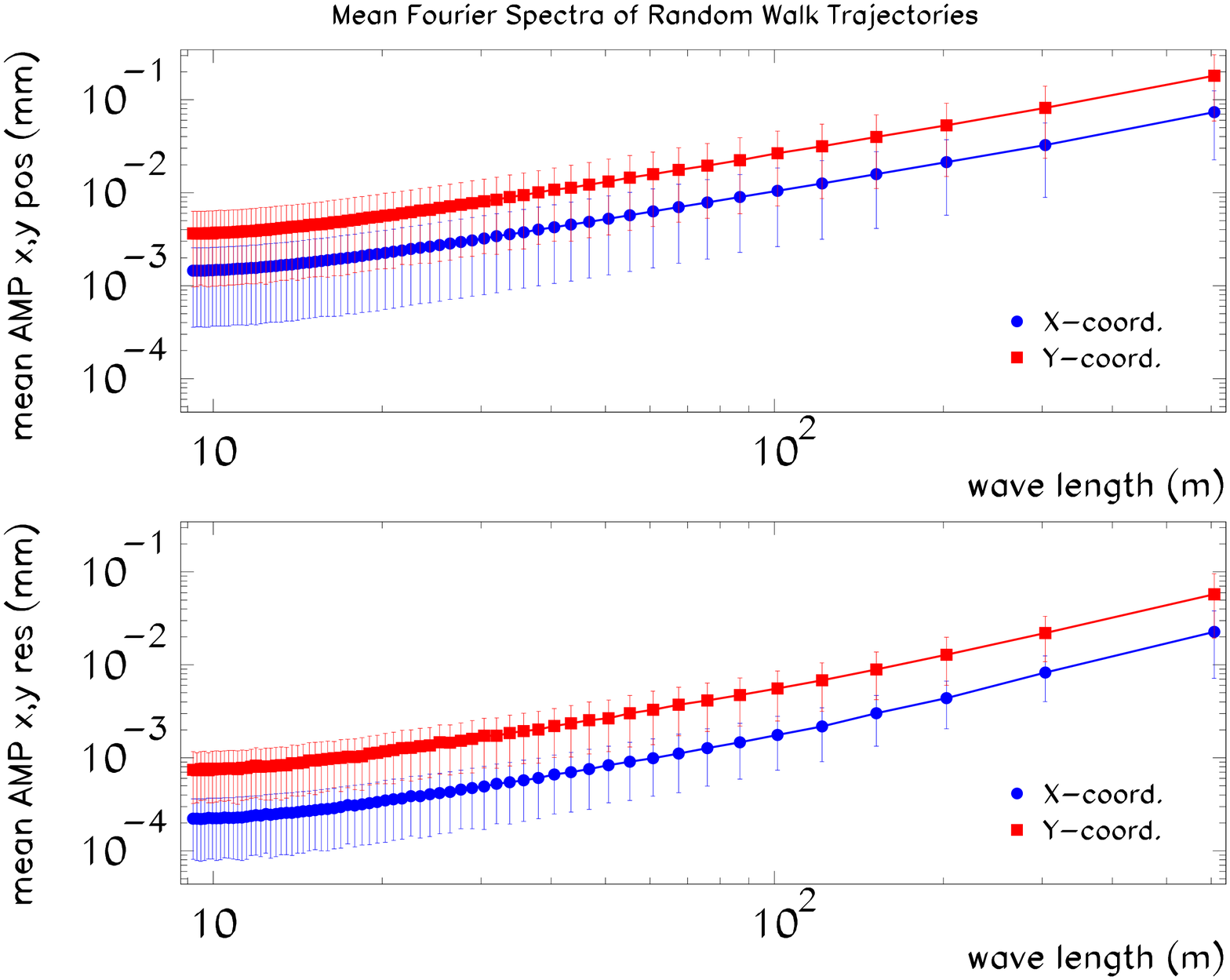}
\end{tabular}
\end{center}
\caption[Fast Fourier Transformation of Random Walk Alignment Trajectory]
{Fast Fourier Transformation (FFT) of random walk alignment trajectories.
Mean value for position amplitude (upper plot) and residua amplitude (lower plot)
from several Monte Carlo generations is presented.}
\label{FFT-TRAJECTORY}
\end{figure}

In order to study the spectra of the alignment trajectories 
obtained form the LiCAS Random Walk model the Fast Fourier
Transformation (FFT) was performed.
In figure \ref{FFT-TRAJECTORY} the mean values of the position and residua
amplitude for several Monte Carlo generations is presented
for $600\,m$ tunnel section. The spectra are dominated by the long wave length
components reaching the amplitude of about $50\,\mu m$ at $600\,m$.
In the future more realistic model should also include the white noise
component from the short distance 'stake out' measurements.

%------------------------------------------------------------------------------
%------------------------------------------------------------------------------

% ****************************************************************************
% BIBLIOGRAPHY AREA
% ****************************************************************************

\begin{footnotesize}
% IF YOU DO NOT USE BIBTEX, USE THE FOLLOWING SAMPLE SCHEME FOR THE REFERENCES
% ----------------------------------------------------------------------------

\end{footnotesize}

% ****************************************************************************
% END OF BIBLIOGRAPHY AREA
% ****************************************************************************

\end{document}